\documentclass[a4paper,11pt]{article}

\usepackage{authblk}
\usepackage{amsmath}
\usepackage{subfig}
\usepackage{hyperref}
\usepackage{xcolor}
\usepackage{fullpage}
\usepackage{todonotes}

\title{Analytical assessment of workers' safety concerning direct and indirect ways of getting infected by dangerous pathogen}

\author[1,*]{Krzysztof Domino}

\affil[1]{{Institute of Theoretical and Applied Informatics, Polish Academy of Sciences}, {Baltycka~5}, {44-100}, {Gliwcie}, {Poland}}

\affil[*]{kdomino@iitis.pl}
\author[1]{Arkadiusz Sochan}
\author[1]{Jarosław Adam Miszczak}

\begin{document}

\maketitle

\begin{abstract}
Developing safety policies to protect large groups of individuals working in indoor environments from disease spread is an important and challenging task. To address this issue, we investigate the scenario of workers becoming infected by a dangerous airborne pathogen in a near-real-life industrial environment. We present a simple analytical model based on observations made during the recent COVID-19 pandemic and business expectations concerning worker protection. The model can be adapted to address other epidemic or non-epidemic threats, including hazardous vapors from industrial processes.
In the presented model, we consider both direct and indirect modes of infection. Direct infection occurs through direct contact with an infected individual, while indirect infection results from contact with a contaminated environment, including airborne pathogens in enclosed spaces or contaminated surfaces. Our analysis utilizes a simplified droplet/aerosol diffusion model, validated by droplet spread simulations.
This model can be easily applied to new scenarios and has modest computational requirements compared to full simulations. Thus, it can be implemented within an automated protection ecosystem in an industrial setting, where rapid assessment of potential danger is required, and calculations must be performed almost in real-time. We validate general research findings on disease spread using a simple agent-based model. Based on our results, we outline a set of countermeasures for infection prevention, which could serve as the foundation for a prevention policy suited to industrial scenarios.
% these are: first, filter out workers who are coughing, reduce workers' mobility and decrease the density of workers in particular compartments, later additional counter-measures such as surface disinfection can be applied. 
\end{abstract}

%\begin{highlights}
%\item Dangerous pathogen spreading in close to real life factory environment has been analysed.
%\item Easy computable simple analytical model has been elaborated and confirmed with droplets' spreading simulations.
%\item Three ways of infection: direct inhalation, indirect inhalation and indirect contact has been analysed.
%\item General countermeasures have been drawn from the simple analytical model.
%\end{highlights}

\section*{Keyword}
industrial safety, direct and indirect infection, droplets spreading,  factory workplace, COVID-19 pandemic

%%%%%%%%%%%%%%%%%%%%%%%%%%%%%%%%%%%%%%%%%%%%%%%%%%
\section{Introduction}
%%%%%%%%%%%%%%%%%%%%%%%%%%%%%%%%%%%%%%%%%%%%%%%%%%

% It is a vital issue, important from the point of view of supply chain maintenance during the epidemics.

The motivation for the presented work stems from the challenges of maintaining supply chain sustainability from the perspective of a single industrial facility during extraordinary situations, such as epidemics. During the recent COVID-19 pandemic, many employers switched to remote work to reduce physical contact between employees and limit pathogen transmission. However, this cannot be a long-term solution, as many jobs require the worker’s physical presence and cannot be performed remotely. Some of these jobs take place outdoors, where the probability of infection is low; in most cases, the risk of infection outdoors is orders of magnitude lower than indoors~\cite{rowe2021simple}. Other jobs, such as those in a factory, must be carried out indoors, where workers are more vulnerable to infections. Nevertheless, many of these factory jobs are considered essential~\cite{milligan2021impact} for sustaining the supply chain.

From the perspective of factory operations, in~\cite{chen2020assessing}, various factors that influenced factory functionality and supply chain robustness during the COVID-19 pandemic were discussed. One of these factors was worker health risks; if a significant number of workers become infected, a factory may need to shut down. In~\cite{li2020intelligent}, an intelligent manufacturing system was proposed to create a safer working environment during the COVID-19 pandemic. This system uses automated manufacturing assets, monitored by networked sensors, to reduce worker contact and, consequently, the probability of infection.

However, developing such systems requires the capability to assess infection probability in real time. This probability can be estimated using a droplet-spreading model based on numerical simulations of the diffusion equation~\cite{vuorinen2020modelling}. However, this approach demands significant time and computational resources, making it potentially unsuitable for edge computing applications in a smart factory setting.

Naturally, infection spread can be analyzed using agent-based models. For example, in~\cite{castro2021multi}, a multi-agent model was used to analyze the spread of COVID-19 infections across various urban spaces, including schools, hospitals, commercial areas, industry, and other common areas. Our model is derived from this approach but is specifically focused on the airborne spread of infectious pathogens in indoor environments, with particular emphasis on industrial facilities.
Technically, we refer to~\cite{karia2020covid} and consider two modes of infection: direct and indirect. The direct mode primarily involves pathogen transmission directly between agents via respiratory droplets produced while breathing, talking, coughing, or sneezing by an infected individual and through body-to-body contact. In contrast, the indirect mode mainly involves contact with contaminated indoor environments~\cite{cai2020indirect, fadaei2021ventilation, bourouiba2020turbulent, morawska2020airborne}, such as air within enclosed spaces, surfaces, tools, utensils, and computer equipment.

To address this problem from an epidemiological perspective, we acknowledge the debate regarding the roles of aerosol versus droplet transmission~\cite{meselson2020droplets} and the spatial range of transmission. Generally, droplets (large particles) remain airborne only in close proximity to the infected individual, primarily leading to direct transmission. The significance of aerosols, which can travel longer distances and remain infectious for extended periods, was recognized later in the pandemic. In our approach, aerosols may contribute to indirect transmission to some extent, as small droplets can travel far, remain suspended in the air for a prolonged time, and carry an active viral load.

According to~\cite{ferretti2020quantifying}, direct transmission is the most probable mode of infection, accounting for approximately $40 \%$ of cases from symptomatic individuals and $50 \%$ from asymptomatic individuals in the case of COVID-19. By comparison, indirect transmission is responsible for an order of magnitude fewer cases than direct transmission. However, the proportion can vary significantly depending on specific circumstances, as there is strong evidence supporting airborne COVID-19 transmission~\cite{greenhalgh2021ten}. Factors such as ventilation rate, directional airflow, and specific activities of infected individuals (e.g., singing) should be considered in assessing transmission risk~\cite{greenhalgh2021ten}.

To enhance industrial safety, we plan to take a more detailed approach to analyzing industrial facilities, where workers frequently share tools, work surfaces, and other equipment. Therefore, we place additional emphasis on indirect contact with contaminated tools, workstations, and shared areas, as such sharing is common practice in factories.
In this context, we divide indirect transmission into two subcategories: contact with \emph{contaminated environmental air} and contact with \emph{contaminated surfaces}.

In detail, we propose a model of infection spread based on three channels of contamination, as developed in~\cite{domino2022will}. Accordingly, we assume the following modes (or channels) of infection spread:
\begin{enumerate}
	\item direct transmission via inhalation of droplets in the \emph{proximity of an infected individual}. For steady breathing, the range of this channel is expected to be approximately $1.5$-$2$m~\cite{wei2016airborne}; however, for loud speech or coughing, the range may be extended~\cite{wei2015enhanced};
	\item indirect transmission via inhalation of small droplets and aerosols from the \emph{contaminated environmental air} in the same compartment as the infected individual. This transmission mechanism is commonly observed in indoor settings, such as workplaces~\cite{greenhalgh2021ten};
	\item indirect transmission via contact with \emph{contaminated surfaces} (e.g., working place or tools that an infected individual has used). This mode of transmission is particularly relevant in industrial facilities, where shared workspaces and equipment are common. 
\end{enumerate}
It is important to note that we consider only inhalation as a means of human-to-human transmission. Other forms of contact, such as handshaking, can be easily eliminated in the factory environment. However, handshaking can be modeled similarly to contact with contaminated surfaces.

The paper is organized as follows. In Section~\ref{sec::analitical_model}, we present the analytical model of droplet spreading based on a simplified diffusion approach. In Section~\ref{sec::case_stud}, we assess infection probability given the analytical model. In Section~\ref{sec::validation}, we validate the model based on droplet spreading simulations. In Section~\ref{sec:abm-results}, we discuss the results of the agent-based simulation of the factory. Finally, in Section~\ref{sec::conclusions}, we explore the implications of the results and provide concluding remarks. 

%%%%%%%%%%%%%%%%%%%%%%%%%%%%%%%%%%%%%%%%%%%%%%%%%%
\section{Analytical model of droplet spreading}\label{sec::analitical_model}
%%%%%%%%%%%%%%%%%%%%%%%%%%%%%%%%%%%%%%%%%%%%%%%%%%

In this section, we model the various modes of pathogen transmission — namely, direct and indirect — as part of the emission and contamination components of our model.
%%%%%%%%%%%%%%%%%%%%%%%%%%%%%%%%%%%%%%%%%%%%%%%%%%
\subsection{Direct transmission – via droplets in close proximity to an infected individual}\label{sec::airborn}
%%%%%%%%%%%%%%%%%%%%%%%%%%%%%%%%%%%%%%%%%%%%%%%%%%

First, we model the droplet density in the \textit{proximity} of \textit{an infected individual}. Following~\cite{vuorinen2020modelling}, we apply the diffusion equation to represent the droplet density, $c(t, x,y,z)$, in the air at a given time $t$ and spatial coordinates $x, y, z$. We use a simplified diffusion approach, where the modeled diffusion accounts for both actual diffusion and convection in the air. Additionally, we assume spherical symmetry and use spherical coordinates.
The rationale for using spherical coordinates (as opposed to the cylindrical coordinates used in~\cite{vuorinen2020modelling}) is that, in industrial facilities, a single floor level cannot always be assumed. 

Following Eq.~$(1)$ in ~\cite{vuorinen2020modelling}, the droplet density in the air, $c(t, r)$, satisfies the following general diffusion equation:
\begin{equation}\label{eq::dif_E}
\frac{\partial c(t,r)}{\partial t} = D \Delta c(t,r) + S(r,t) - \frac{c(t,r)}{\tau},
\end{equation} 
where $\Delta$ is the Laplace operator:
\begin{equation}
\Delta = \partial^2_r + \frac{2}{r} \partial_r,
\end{equation}
and the spherical symmetry yields $\partial^2 \theta  = \partial \theta = \partial^2 \psi  = \partial \psi = 0$.
The parameter $D$ represents the diffusion coefficient, accounting for both actual diffusion and convection in the air. The parameter $S(r,t)$ denotes the droplet source rate, and $\tau$ is the decay constant.

The droplet density decays with a half-life $T_{1/2} = 0.693 \tau$, as droplets are removed through processes such as drying, sedimentation, or ventilation. In~\cite{vuorinen2020modelling}, the decay constant $\tau$ was assumed to be on the order of $100$s. However, determining the exact value of $\tau$ is not straightforward and requires sensitivity analysis and validation.
 Nevertheless, $\tau$ is relatively small compared to the time required for an agent to complete a particular industrial task.  This assumption allows for a significant simplification: we can approximate Eq.~\eqref{eq::dif_E} by treating it as time-independent (i.e., setting $\partial c / \partial t = 0$), which makes it solvable analytically. (While performing such tasks, the agent is assumed to remain in a relatively fixed location.)

Following~\cite{vuorinen2020modelling}, we assume that, in unit time, an individual releases $s$ units of contaminated air into the control volume  $V_{\text{control}} = 1 m^3$ (in our case, a sphere with a radius of $r_0 = 0.62$m). Within this volume, air and droplets mix uniformly, (i.e., droplets are distributed evenly for $r < r_0$), while diffusion and convection occur beyond this control volume. Then, for $r \geq r_0$, the source rate in Eq.~\eqref{eq::dif_E} is:
\begin{equation}
S(r) = \begin{cases} s / V_{\text{control}} &\ \  r = r_0 \\
0 &\ \  r > r_0
\end{cases}.
\end{equation}
We introduce the parameter $d_\text{max}$, which limits the validity of the model (for direct transmission) within a spherical distance from the source, leading to the boundary condition $c(d_\text{max}) = 0$.
The droplet density at a distance $r$ from the source is then described by the following formula:
\begin{equation}
\label{eq::concentration}
\hat{c}(r) = \begin{cases}
C &\text{ for } r \leq r_0 \\
c(r) &\text{ for } d_{\max} > r > r_0  \\
0 &\text{ for } r \geq d_{\max}
\end{cases}.
\end{equation}
For $r_0 \leq r \leq d_{\max}$ we can solve Eq.~\eqref{eq::dif_E} analytically with conditions:
\begin{enumerate}
 \item $\partial c / \partial t = 0$, 
 \item $c(r_0) = C$, $S(r) = s \cdot \delta(r-r_0)$,
 \item $c( d_{\max}) = 0$.
\end{enumerate}
The resulting expression is:
\begin{equation}\label{eq::cr}
c(r) = \frac{C  r_0 \text{csch}\left(\frac{d_{\max}-r_0}{\sqrt{D \tau}}\right) \cdot \sinh\left(\frac{d_{\max}-r}{\sqrt{D \tau}}  \right) }{r} \ \ \text{droplets}/ \text{m}^3.
\end{equation}
The parameters are listed below:
\begin{enumerate}
	\item $D$ -- the diffusion coefficient, reflecting the actual diffusion and convection of droplets in the air. In~\cite{vuorinen2020modelling}, a value of $D = 0.05 $m$^2$/s was used.
	\item  $\tau$ -- the decay constant, representing the rate of droplet removal due to processes like drying and ventilation. According to~\cite{vuorinen2020modelling}, $\tau$ is expected to be on the order of $100$s.
	\item $d_{\max}$ -- the limit of the model validity. Here, we use $d_{\max} = 2.6$m. For steady breathing, droplets travel an average distance of $1.5$-$2$m from the source individual~\cite{wei2016airborne}; however, for loud speech or coughing, droplets can travel significantly farther~\cite{wei2015enhanced}. 
	\item $C$ -- the model's constant that depends on the decay constant, source rate, and other parameters of the model.
\end{enumerate}
To compute the constant $C$, we assume the total number of droplets to be $s \tau$. Assuming spherical symmetry, we have:
\begin{equation}
\int_0^{\infty} \hat{c}(r) 4 \pi r^2 dr = s \tau,
\label{eq::Ccomp}
\end{equation}
we compute constant $C$ in terms of $s, \tau, r_0, d_{\max}$. 

The parameter $s$, which represents the rate of droplets produced by an individual, requires further discussion and validation.
Initially, we refer to the works of~\cite{vuorinen2020modelling, asadi2019aerosol}, which suggest the following values for this parameter:
\begin{itemize}
\item $s = 1.5 $s$^{-1}$ for breathing individual,
\item $s = 5 $s$^{-1}$ for speaking individual, 
\item $s = 66 $s$^{-1}$ for coughing/sneezing individual.
\end{itemize}
In the case of wearing a mask, the source intensity is reduced by approximately $30\% - 40 \%$ \cite{leung2020respiratory}. (For coughing/sneezing with a mask, this results in $s \approx 40$ s$^{-1}$). 

\begin{figure}\includegraphics[width=1.0\textwidth]{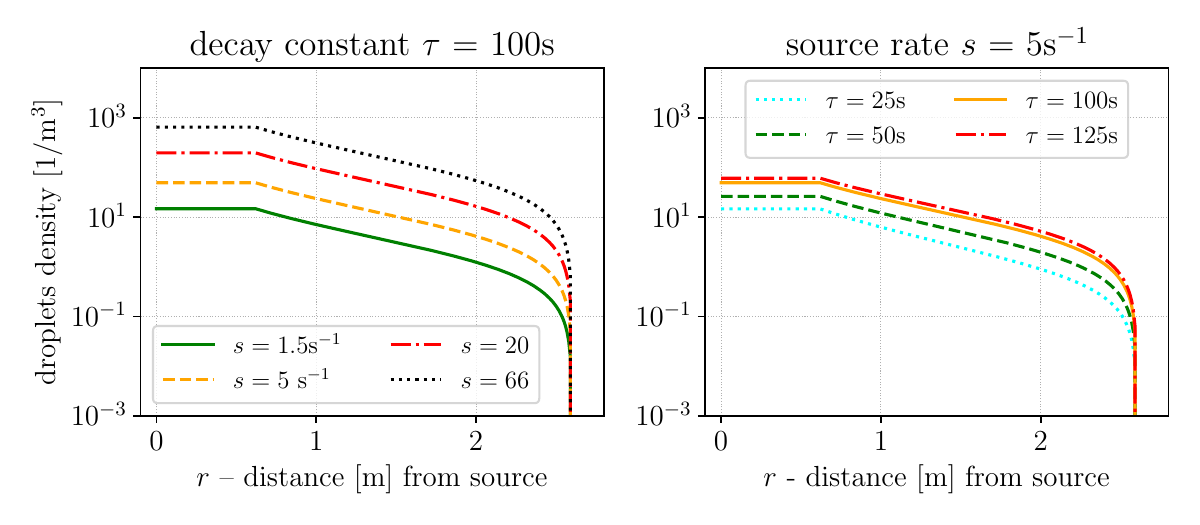}
	\caption{Droplet density in the proximity of an infected individual, swap over source rate and decay constant parameters. We use decay constant $\tau = 100$s and source rates $s$ in the range of $1.5$ - $66$s$^{-1}$ in the left panel, and source rate $s = 5$s$^{-1}$ and decay constant in the range of $25$ - $125$s in the right panel.}\label{fig::sol_EQ}
\end{figure}

Returning to the model, the droplet density is given by:
\begin{equation}\label{eq::crs}
\hat{c}(r, s) = \begin{cases}
C(s, \tau, r_0, d_{\max}) &\text{ for } r \leq r_0 \\
 C(s, \tau, r_0, d_{\max})\frac{r_0}{r} \text{csch}\left(\frac{d_{\max}-r_0}{\sqrt{D \tau}}\right) \cdot \text{sinh}\left(\frac{d_{\max}-r}{\sqrt{D \tau}}  \right)  &\text{ for } d_{\max} > r > r_0  \\
0 &\text{ for } r \geq d_{\max}
\end{cases},
\end{equation}
obviously $\hat{c}$ depends also on $\tau, r_0, d_{\max}$. For particular values of droplet's density see Fig.~\ref{fig::sol_EQ}.

\begin{figure}
    \centering
	\includegraphics[width=0.6\textwidth]{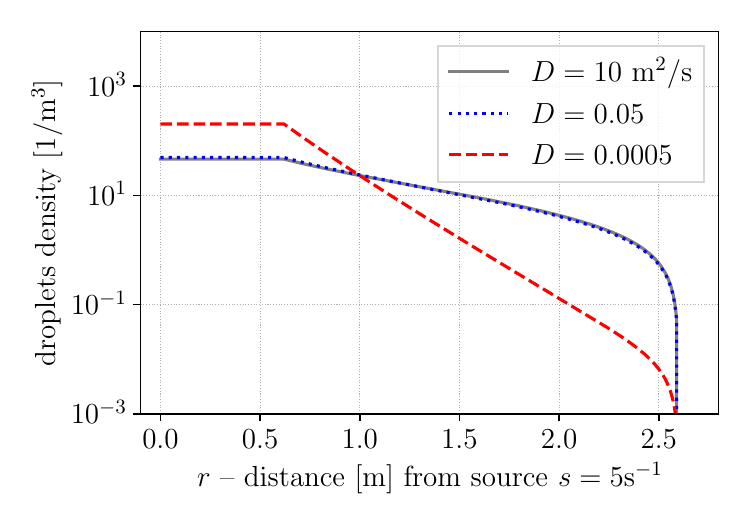}
	\caption{Droplets density at distance $r$ to the source with the rate $s = 5$s$^{-1}$, sensitivity analysis for various diffusion constant $D$ values used in Eqs.~\eqref{eq::dif_E}~\eqref{eq::crs} (we use decay constant $\tau = 100$s). Observe similar droplet density values for the wide range of $D$ values.}\label{fig::sensitivity}
\end{figure}

In the final part of this subsection, we note that the value of the parameter $D$, which reflects the diffusion and convection of air, is heuristic in nature. The specific value used in our model is based on the heuristic choice made in~\cite{vuorinen2020modelling}. 
However, if $D$ is in some reasonable range, the model is not highly sensitive to its particular value, see Fig.~\ref{fig::sensitivity}. (We observe from Eqs.~\eqref{eq::cr} and \eqref{eq::Ccomp} that changes in $D$ can be approximately counterbalanced by adjustments in $C$.) 
We also acknowledge that our approach is an approximation, as it does not account for the size of the droplets.

We also note, that in reality, droplets of varying sizes travel different distances, but incorporating such detailed considerations would require much more computationally expensive simulations, which would compromise the simplicity of the model.

%%%%%%%%%%%%%%%%%%%%%%%%%%%%%%%%%%%%%%%%%%%%%%%%%%%%%%%%%%%%%%%%%%%%%
\subsection{Indirect transmission – contaminated environmental air}\label{sec::indirect_air}
%%%%%%%%%%%%%%%%%%%%%%%%%%%%%%%%%%%%%%%%%%%%%%%%%%%%%%%%%%%%%%%%%%%%%

To model pathogen spread within a compartment, particularly at distances farther from the source, we refer to indirect transmission via contaminated indoor air~\cite{cai2020indirect}. At longer distances, the transmission relies primarily on small droplets and aerosols~\cite{meselson2020droplets}, while the spread near the source is more influenced by larger droplets. The dynamics of this indirect transmission are complex and depend heavily on the compartment's spatial layout and airflow characteristics. However, since this mode of transmission is not expected to be dominant, we simplify the analysis by applying a mean approximation. Specifically, we assume that the droplets (or aerosol particles) that contaminate the air are distributed uniformly throughout the compartment.

Assuming there are $S' \in \{s_1, s_2, \ldots\}$ sources (infected individuals) within a compartment of volume $V$, and following the approach in~\cite{vuorinen2020modelling} (or using discussion analogous to Eq.~\eqref{eq::Ccomp}), we approximate the stationary state of the droplet (or aerosol) density in the compartment as: 
\begin{equation}
	\label{eq::cbackgropud}
	c_{\text{b}}(s_1, \ldots) = \frac{\tau' \sum_{s \in S'} s}{V} \ \  \text{droplets}/ \text{m}^3.
\end{equation}
We recognize that indirect transmission, as discussed in this subsection, is particularly relevant when mediated by small droplets (or aerosol particles) that can remain suspended in the air for longer periods. In indirect transmission, aerosols may play a more dominant role. The decay constant denoted by $\tau'$, differs from the decay constant $\tau$ used for the direct transmission model in Eq.~\eqref{eq::cr}, and is generally higher. The precise determination of $\tau'$ requires careful consideration and should ideally be validated through simulations. 

\begin{figure}[ht]
    \centering
	\includegraphics[width=1.0\textwidth]{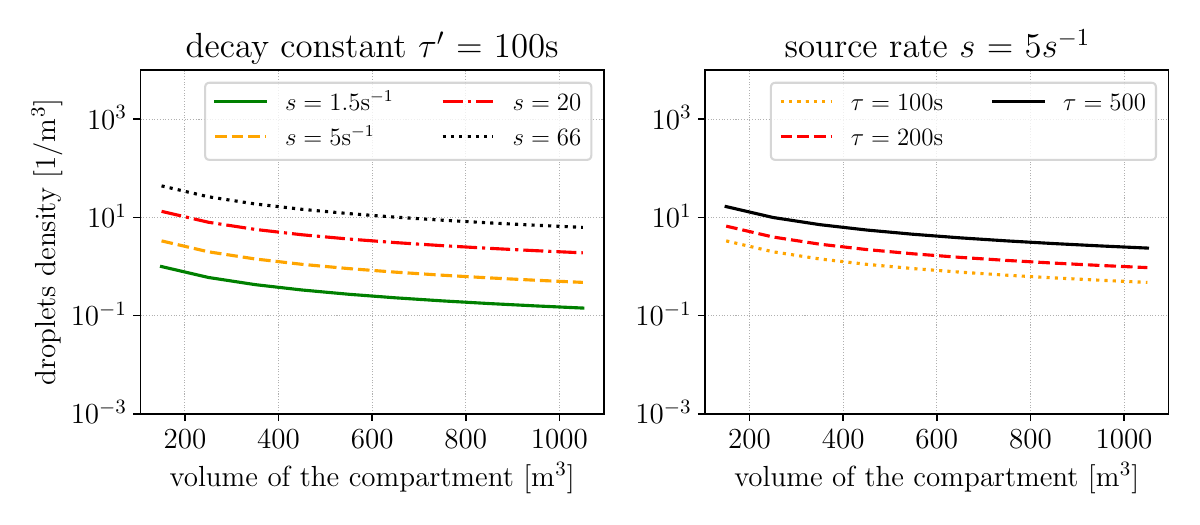}
	\caption{Droplets density, swap over parameters for a density of droplets in \textit{contaminated environmental air}, with one infected individual of source rate $s$.}\label{fig::background_droplets}
\end{figure}

In Fig.~\ref{fig::background_droplets} we present the density of droplets given particular ranges of values of parameters for indirect transmission via environmental air. In this particular parameter range, densities are considerably lower than those resulting from direct transmission in Fig.~\ref{fig::sol_EQ}.

%%%%%%%%%%%%%%%%%%%%%%%%%%%%%%%%%%%%%%%%%%%%%%%%%%
\subsection{Indirect transmission -- contaminated surface}\label{sec::indirect_sedimentation}
%%%%%%%%%%%%%%%%%%%%%%%%%%%%%%%%%%%%%%%%%%%%%%%%%%

We expect that, for factory workers, contact with contaminated workplaces or tools can also be a significant source of infection. Surfaces are typically contaminated by the sedimentation of droplets exhaled by an infected individual or from generally contaminated air. Here, we will focus on contamination caused by individuals, as this route is expected to be more relevant.
This approach aligns with the modeling in~\cite{vuorinen2020modelling} (and Fig. 2.1D therein), where droplets of diameter $80 \mu $m or greater can sediment before evaporation. Therefore, to model the surface contamination, we are only interested in large droplets. Referring to~\cite{chao2009characterization} (and Tab. $1$ and Fig. $3$ therein) it can be concluded that in exhaled air, approximately $w = 10 \%$ of the droplets have a diameter of $80 \mu $m or greater.

As the accurate model of droplet sedimentation is complex, following Section~\ref{sec::indirect_air}, we propose an approximation of uniform surface contamination. Under this approximation, the source rate -- i.e., the number of droplets sedimented on a unit area per unit time is given by
\begin{equation}
S_{s} = \frac{w s}{A},
\end{equation}
where $A$ is the area of the surface that is contaminated. The density of droplets on the surface (per unit area) can be modeled by the following decay equation:
\begin{equation}\label{eq::c_surface}
	\frac{\partial c_{s}(t)}{\partial t} = S_{s} - \frac{c_{s}(t)}{\tau_{s}},
\end{equation}
where $\tau_{s}$ is the decay constant for the surface, which differs significantly from the decay constant in the air (both $\tau$ and $\tau'$). This is because droplets in the air can evaporate, be removed by ventilation, etc. To illustrate this difference, consider the example provided in~\cite{van2020aerosol}, where half-life times $T_{s, 1/2} = \tau_s \ln(2)$ for the COVID-19 virus on different surfaces were recorded: $5$ hours on stainless-steel and $7$ hours on plastic surfaces (both materials commonly found in industrial settings). These timescales, as mentioned in~\cite{van2020aerosol}, are of the same order of magnitude as the typical length of a working shift. 
In conclusion, for surface contamination by pathogens similar to COVID-19, we can neglect the decay factor in Eq.~\eqref{eq::c_surface}, yielding:
\begin{equation}\label{eq::cs_surface}
c_s(s) = \frac{w s}{A} t,
\end{equation}
where $t$ is the time of the droplet's sedimentation process.

For the possible decay mechanism, we can consider disinfection. The disinfection of the surface will reduce $c_s$ by a certain percentage, as discussed in~\cite{kampf2020persistence}, particularly in Table III.

\section{Probability of infection}\label{sec::case_stud}

As our research focuses on airborne pathogens, we equate the spreading of droplets (or aerosol particles, which are considered here as very small droplets) with the transmission of pathogens. In this context, infection can occur either through the inhalation of infected droplets or aerosols (which may spread directly or indirectly) or through contact with a surface contaminated by droplets that have already sedimented (indirect transmission).

\subsection{Inhalation of droplets}\label{sec::droplets_inhaled}

In this subsection, we calculate the probability of infection through inhalation, regardless of the droplet spreading mechanism considered (either direct transmission from the proximity of an infected individual or indirect transmission from contaminated environmental air). Let $c$ represent the density of droplets per unit volume, as given by Eq.~\eqref{eq::crs} (in the case of direct droplet source from proximity of infected individual), or by Eq.~\eqref{eq::cbackgropud} (for the case of indirect droplet transmission from contaminated environmental air). The number of droplets inhaled over a time period $t$ (assuming constant $c$ from Eq.~\eqref{eq::crs} or Eq.~\eqref{eq::cbackgropud}) is:
\begin{equation}\label{eq::n_droplets}
N(t) = Q c t,
\end{equation}
 where the breathing rate $Q$ is the model parameter. The breathing rate value is influenced by the attributes of the individual, such as physical effort, job type, and the tasks being performed. To determine a particular value for $Q$, we refer to~\cite{vuorinen2020modelling}, where $Q$ was modeled probabilistically with a uniform distribution in the range of $8 - 30$dm$^3/$min, or using its mean value of $19$dm$^3$/min. In this section, we adopt the latter approach; however, the model can easily be adapted to the probabilistic one. It is worth noting that the value of $19$dm$^3$/min may seem relatively high, especially when compared to medical measurements (e.g., respiratory machines). However, we believe this value is more representative of factory workers performing physically demanding tasks.

\begin{figure}	\includegraphics[width=1.0\textwidth]{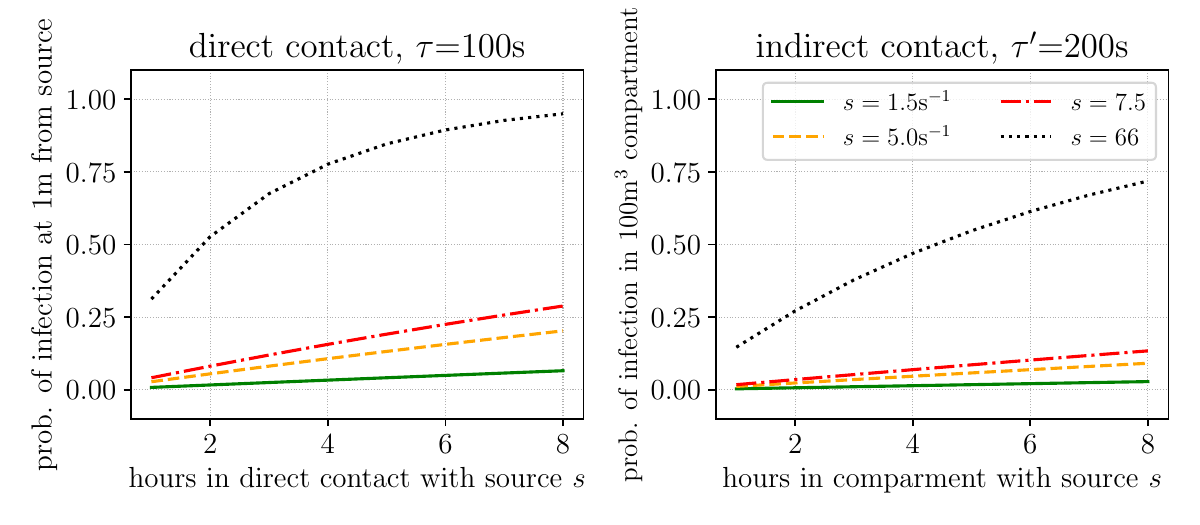}
\caption{Probability of infection in \textit{a direct} mode from \textit{proximity of an infected individual} (left panel), or in \textit{indirect} mode from \textit{contaminated environmental air} (right panel). We assume various  source rates $s$ (from $1.5$s$^{-1}$ to $66$s$^{-1}$ ). We assume $N_b= 100$ in Eq.~\eqref{eq::probability_air}  in all cases. In the right panel, we use compartment volume of $V = 100$m$^3$.}\label{fig::probs}
\end{figure}

To compute the probability of infection via inhalation, it is necessary to determine the number of droplets, $N_b$, that are exhausted by the infected individual and subsequently inhaled by the healthy individual. This number should correspond to the threshold probability of infection, for example, $a = 10 \%$.
Unfortunately, $N_b$ is not a universal parameter, as it depends on various factors such as pathogen type, the infected individual's condition, immunity, etc. According to research on COVID-19 from 2020~\cite{vuorinen2020modelling}, a significant number of individuals become infected when they inhale between $10 \leq N_b \leq 1000$ droplets exhaled by an infected individual. Referring to Tab $3$ in~\cite{vuorinen2020modelling} and additional data from in~\cite{lu2020covid}, it can be concluded that, on average, inhaling $N_b = 100$ droplets results in an infection probability of approximately $a=10\%$. 
Thus, the probability of infection can be modeled as:
\begin{equation}\label{eq::probability_air}
p_d(t) = 1-(1-a)^{\frac{N( t)}{N_b}},
\end{equation}
where $N(t)$ is the number of droplets inhaled over time $t$. Note  that by Eq.~\eqref{eq::n_droplets} $N$ is the function of $Q$ and parameters of $c$ from  Eq.~\eqref{eq::crs} or Eq.~\eqref{eq::cbackgropud}, e.g. the source rate. 

The infection probabilities computed in this way are presented in Fig.~\ref{fig::probs}  (\textit{direct transmission} from \textit{proximity of infected individual} -- left panel, \textit{indirect transmission} from \textit{contaminated environmental air} -- right panel). 
From this analysis, we can conclude that the probability of infection through indirect transmission (droplets from the background), even in a relatively small industrial compartment with a volume of $100$m$^3$, is lower than the probability of infection due to direct contact. Following Eq.~\eqref{eq::cbackgropud} we expect the relative importance of indirect transmission to decrease further in larger industrial compartments (e.g., those with volumes in the range of several $1000$m$^3$). This observation aligns with the findings in~\cite{ferretti2020quantifying}, where direct transmission was identified as the most probable route of infection. Therefore, we will not consider the \textit{indirect} transmission route via \textit{contaminated environmental air} in the agent-based simulations.

\begin{figure}
\centering\includegraphics[width=0.6\textwidth]{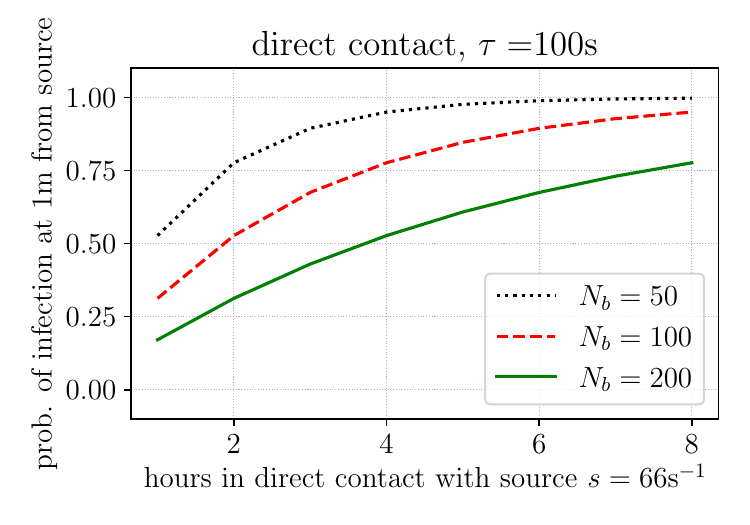}
\caption{Probability of infection from direct contact with an infected individual of high source rate ($s = 66$s$^{-1}$), for various immunity factors $N_b$ in Eq.~\eqref{eq::probability_air}. Namely: $N_b = 50$ - low immunity, $N_b = 100$ - normal immunity, $N_b = 200$ - high immunity (e.g. vaccinated).}\label{fig::probsNb}
\end{figure}

The sensitivity analysis for various $N_b$ is presented in~Fig.~\ref{fig::probsNb}. In this analysis, $N_b = 100$ represents the baseline scenario, $N_b = 50$ reflects an agent with low immunity or a highly infectious pathogen, and $N_b = 200$ corresponds to an individual with higher immunity (e.g., vaccinated). The key takeaway is that prolonged contact between an agent with low immunity and an agent with a high source rate (e.g., coughing) results in nearly certain infection. This observation serves as a straightforward consistency check for our model.

%%%%%%%%%%%%%%%%%%%%%%%%%%%%%%%%%%%%%%%%%%%%%%%%%%
\subsection{Infection by contact with a contaminated surface}\label{sec::contact}
 %%%%%%%%%%%%%%%%%%%%%%%%%%%%%%%%%%%%%%%%%%%%%%%%%%
In this subsection, we consider the \textit{indirect transmission} of infection via \textit{contaminated surface}. This form of transmission is estimated to account for approximately $10\%$ of COVID-19 infections~\cite{ferretti2020quantifying}. As we have not found an established model for calculating the probability of infection through \textit{contaminated surfaces}, we propose the following simplified model.
Let $c_s$ be the number of droplets exhaled by an infected individual onto a unit area of a surface. Once a sufficient number of droplets accumulate, the surface can be considered contaminated. To determine a contamination threshold for surfaces, we refer to ~\cite{wilson2021modeling}, where a threshold of $1$ genome copy (gc) per /cm$^2$ was used to indicate surface contamination. Furthermore, as noted in~\cite{wilson2021modeling}, if a surface is contaminated, the probability of a healthy individual contracting an infection through hand-to-fomite and subsequent hand-to-face (mouth, eyes, or nose) contact is approximately $1:10000$ per contact.

To define the model parameters for COVID-19 as the pathogen, we refer to~\cite{wolfel2020virological}, where the mean viral load of the virus is reported as $\text{vir}_{\text{load}} = 7 \times 10^6$ genome copies (gc) per ml, which translates to $7 \times 10^{12}$ gc per m$^3$ of saliva. Droplet diameters, as discussed in~\cite{vuorinen2020modelling} (Fig. 2.1D), range from $1$ to $200, \mu$m, with large droplets that are likely to sediment typically having diameters between $80$ and $200, \mu$m, as indicated by~\cite{chao2009characterization} (Tab. $1$ and Fig. $3$).
In Section~\ref{sec::indirect_sedimentation}, we assume that approximately $w = 10\%$ of all droplets are large. For simplicity, we approximate the mean diameter of these large droplets as $140, \mu$m. This gives a mean volume for a large droplet of approximately $\text{vol}_d = 1.44 \times 10^{-12}$ m$^3$. The average number of viral genomes contained within a large droplet can then be estimated as:
\begin{equation}
\nu \approx \text{vir}_{\text{load}}\text{vol}_d \approx 10 \text{gc / droplet}.
\end{equation}
Let $A$ represent the area on which large droplets settle. The viral density on this area, after a source with droplet production rate $s$ continuously emits droplets over a duration $t$, can be given by (see Eq.~\eqref{eq::cs_surface}):
\begin{equation}
gc = c_s(w,s,A,t) \nu  = \frac{w s t}{A} \nu.
\end{equation}
To exceed the threshold viral density on the surface, defined as $gc_{\text{thres}} = 1$ gc /cm$^2 = 10000$ gc /m$^2$ (above which the surface is considered contaminated), droplets need to settle for a duration equal to or greater than a critical time threshold:
\begin{equation}\label{eq::tthres}
t_{\text{thres}} = \frac{A \cdot  gc_{\text{thres}}}{w s \nu}.
\end{equation}
This time threshold ($t_{\text{thres}}$)  is, however, highly dependent on the area of the workspace ($A$), as illustrated in the following examples:
\begin{itemize}
    \item for $A = 2.25$m$^2$ and a source rate $s = 66$ we find  $t_{\text{thres}} \approx 6$ minutes;
    \item for $A = 2.25$m$^2$ and $s = 5$,  $t_{\text{thres}} \approx 75 $minutes;
    \item for $A = 1$m$^2$ and $s = 66$,  $t_{\text{thres}} \approx 2.5$ minutes;
    \item for $A = 1$m$^2$ and $s = 5$, $t_{\text{thres}} \approx 33$ minutes.
\end{itemize}
In conclusion, for small values of $s$ the threshold time $t_{\text{thres}}$ is relatively long. In such scenarios, frequent disinfection of the surface (e.g., once per hour) can effectively reduce the probability of infection from contact with a contaminated surface.
Given the significant variation in $t_{\text{thres}}$ across different scenarios, we will conduct simulations with a variable surface contamination probability. 

To further evaluate the model, recall that when the surface is contaminated, for each hand-to-fomite and hand-to-face contact, the probability of infection is assumed to be $p_{\text{contact}} = 10^{-4}$~\cite{wilson2021modeling}. If such contact occurs with a frequency $f_{\text{contact}}$, the probability of infection is given by:
\begin{equation}\label{eq::prob_surface}
p(t, t_1) = \begin{cases}  t f_{\text{contact}} p_{\text{contact}}  &\text{ if } t_1 > t_{\text{thres}}  \text{ Eq.~\eqref{eq::tthres}}\\ 0 &\text{ otherwise} \end{cases},
\end{equation}
where $t_1$ is the time spent by an infected agent at the particular surface or device prior, and $t$ is the time spent by the healthy agent at the contaminated surface afterward (here, no disinfection is assumed between the presence of agents at the surface).

\begin{figure}
	\centering
	\includegraphics[width=0.49\textwidth]{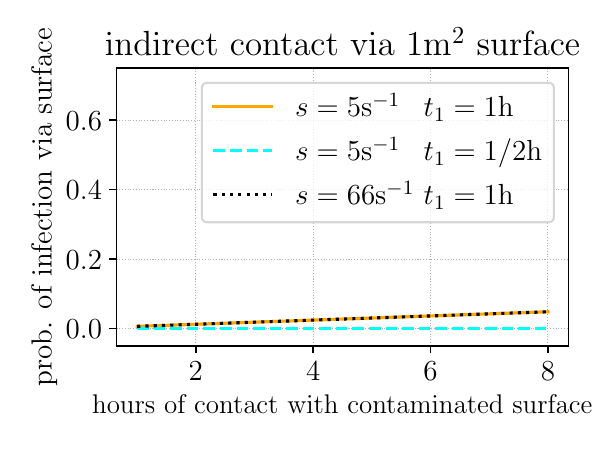}
	\includegraphics[width=0.49\textwidth]{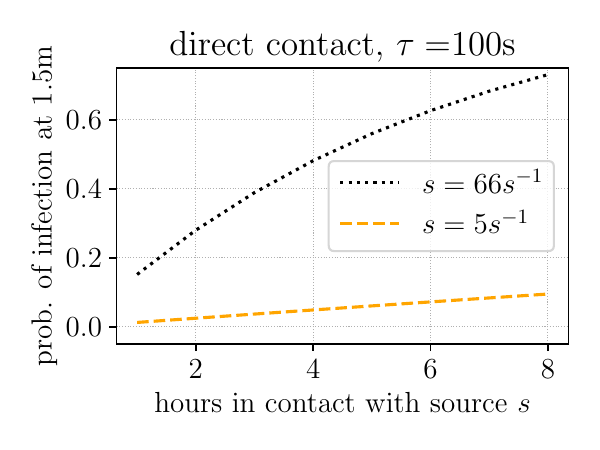}
	\caption{The probability of infection in \textit{indirect} mode from \textit{contact with a contaminated surface} (left panel) compared with the probability of infection in \textit{direct} mode from \textit{proximity of infected individual} (right panel). In the left panel, we use various surface contamination scenarios, namely before the contact the infected agent (the source with rate $s$) stays $t_1$ hours at the surface. We use $p_{\text{contact}} = 10^{-4}$~\cite{wilson2021modeling} for modeling.}\label{fig::prob_surface}
\end{figure}

Infection probabilities in this scenario are depicted in Fig.\ref{fig::prob_surface} — the left panel (where we assume one hand-to-fomite and hand-to-face contact per minute, i.e., $f_{\text{contact}} = 1$ minute$^{-1}$). These probabilities are compared with analogous scenarios of infection through \textit{direct transmission} (see Fig.\ref{fig::prob_surface}, right panel). Although, in most cases, infection via the \textit{direct transmission} is considerably more probable, the possibility of indirect infection through contaminated surfaces cannot be neglected, even in a large-volume industrial hall. For this reason, such \textit{indirect transmission} channels will be analyzed in agent-based simulations, as presented in Section~\ref{sec:abm-results}.

%%%%%%%%%%%%%%%%%%%%%%%%%%%%%%%%%%%%%%%%%%%%%%%%%%
\section{Simulation of droplets spreading -- validation}\label{sec::validation}
%%%%%%%%%%%%%%%%%%%%%%%%%%%%%%%%%%%%%%%%%%%%%%%%%%

The analytical model presented in previous sections is straightforward and computationally efficient. However, it involves numerous assumptions and simplifications regarding droplet spreading that require validation. For this purpose, we have chosen to validate first the model for \emph{direct} droplet spreading in the proximity of the source, as this mode of transmission is the most significant in the context of disease spread. We aim to validate the model through droplet dynamics simulations. It is important to note, however, that such simulations are computationally demanding; the results presented here are from simulations that required 9–10 days of computation on a GPU (GeForce GTX 1080 Ti – 3584 CUDA cores and 11GB memory), depending on the simulation parameters.

We performed simulations of droplet dynamics in a $5\times5\times2.5$ m$^3$ room using the following setup. The Lattice Boltzmann method \cite{kruger2017lattice}, implemented via the \emph{lbmpy} Python module, was used to simulate the motion of labeled air particles. The following parameters were applied: kinematic viscosity of $1.516 \times 10^{-5}$ m$^{2}/$s, maximum physical velocity of 1 m/s, D3Q15 equilibrium moments, a Single Relaxation Time model with a relaxation rate of $1.994$, and the Smagorinsky turbulence model. A proprietary breathing model was employed to induce changes in respiratory air velocity.
For simulating droplet particle movement relative to the labeled air particles, an advection-diffusion approach was used, with a diffusion coefficient of $23.9 \times 10^{-6}$ m$^{2}/$s and a Q27 diffusion direction pattern. The droplet particles initially reside only in the lungs of the infected individual.
Unlike large droplets that settle, as discussed in Section~\ref{sec::indirect_sedimentation}, we approximate small droplets here, neglecting droplet decay (decay is instead approximated by limiting the simulation duration). Specifically, we followed the simulation methodology from~\cite{papadakis2021kyamos} and considered small droplets with a mean diameter of $1 \mu$m. This choice aligns with the massless particle approximation (for particles up to $10 \mu$m in diameter), as outlined in Section $7.3$ of~\cite{vuorinen2020modelling}.
From the simulations, we obtained a unit-less droplet density, which we converted to droplets/m$^3$ to ensure compatibility with the results in Section~\ref{sec::analitical_model}. It is worth noting that our simulations did not account for air movement driven by factors other than breathing. As a result, in real scenarios, the observed spread of droplets may be slightly greater than in these idealized simulation conditions.

We conducted two simulation settings: emit and absorb. Each simulation runs for 20 minutes, with results collected continuously at specified time intervals.
\begin{enumerate}
\item \emph{Emit Setting}: This setting focuses on the spread of droplet density in the vicinity of the infected individual. The emit setting is particularly useful for evaluating the analytical model of droplet density in proximity, as described by Eq.~\eqref{eq::crs}.
\item \emph{Absorb Setting}: This setting simulates the absorption of droplets or aerosols by a healthy individual from the surrounding air. The droplet density in the air was obtained from the emit simulation after a complete 20-minute cycle, with varying distances between the healthy and infected individuals (denoted by $r$). The absorb setting is therefore suitable for assessing the overall infection risk within the space, taking into account the specific location of the infection source.
\end{enumerate}

\begin{figure}[ht]
    \centering
	\includegraphics[width=0.45\textwidth]{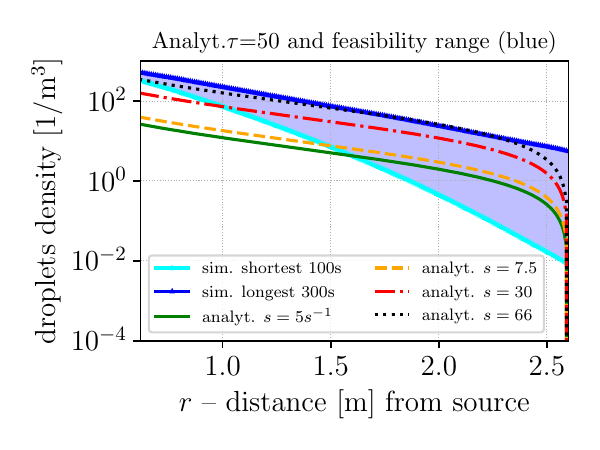}
 \includegraphics[width=0.45\textwidth]{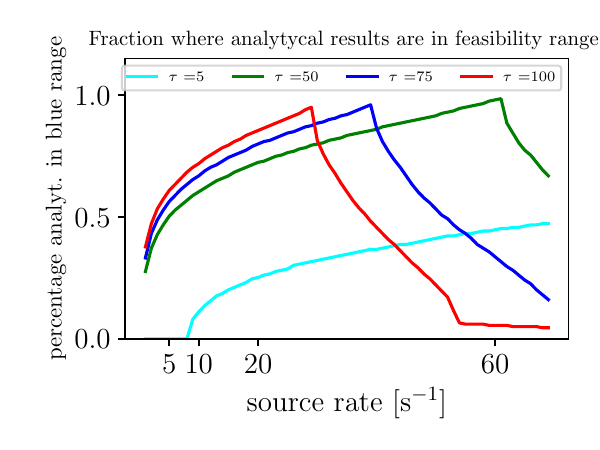}
	\caption{Emit setting. Left panel, assessment of the analytical model of droplets in proximity (Eq.~\eqref{eq::crs}) by the feasibility range computed from simulations (computed for simulation time between $100$s and $300$s) - the blue area (the range on $x$-axis limited by $r_0=0.62$ and $d_{\max} = 2.6$). Right panel, assessment of particular decay constant $\tau$ [s], where the percentage of analytical result in the feasibility range from simulations is depicted; observe, that $\tau=50$s models well high source rate, what is important in our model.}\label{fig::simulations}
\end{figure}

The evaluation of the analytical model using the emit simulation results (after the full $20$-minute cycle) is shown in Fig.~\ref{fig::simulations}. For the simulation output, distance is measured on the $y$-axis as a $1$D slice of $3$D spatial locations in the direction of the mouth (the droplet density values in other directions are similar within the valid region).
As noted, the simulations do not account for droplet decay processes (e.g., evaporation, sedimentation). To address this limitation, we set the simulation time to match the order of magnitude of the decay constant discussed in Section~\ref{sec::analitical_model} and \cite{vuorinen2020modelling}. Specifically, we use simulation times ranging from $100$ to $300$ seconds, where shorter times are intended to model non-stationary states, while longer times extend well beyond the decay half-life in close proximity to the infection source.

From Fig.~\ref{fig::simulations} (left panel), we observe that the analytical model's outputs generally fall within the range of values returned by simulations for scenarios with high source intensity and for distances greater than $1.5$ m when considering lower source intensities. This outcome aligns with expectations, as workers typically do not operate in very close contact.
Looking at the right panel of Fig.\ref{fig::simulations}, there is a reasonable agreement between the simulations and the analytical model across the range of source intensities, with the best match occurring at a decay time constant of $\tau = 50$ s. This value is consistent with reference data from~\cite{vuorinen2020modelling}.
However, from Fig.~\ref{fig::simulations} (left panel), it is evident that for low source rate values ($s$), the analytical model does not align well with simulation results across the entire distance range from the source. Specifically, when the source rate is low, the analytical model performs poorly at approximating simulation results for distances closer than $1.5$ m from the source. In the worst-case scenario, at very close distances (significantly less than $1$ m), the model’s estimates may be up to an order of magnitude lower than those observed in the simulations.
%To overcome this, at least partially, we suggest raising the source rate value for speaking individual from $s=5$ s$^{-1}$ suggested in Section~\ref{sec::analitical_model} to $s=7.5$ s$^{-1}$.
In the analytical model thus far, we applied Eq.~\eqref{eq::crs}, which considers only direct infection from proximity to an infected individual (i.e., the direct pathway). However, in our simulations, the infected individual is situated within a compartment of a specific volume ($62.5$ m$^3$). A more comprehensive analysis should also include the indirect pathway through contaminated environmental air. This indirect transmission route will be addressed in the next simulation setting, discussed in the following paragraph.

\begin{figure}[ht]
    \centering
	\includegraphics[width=0.45\textwidth]{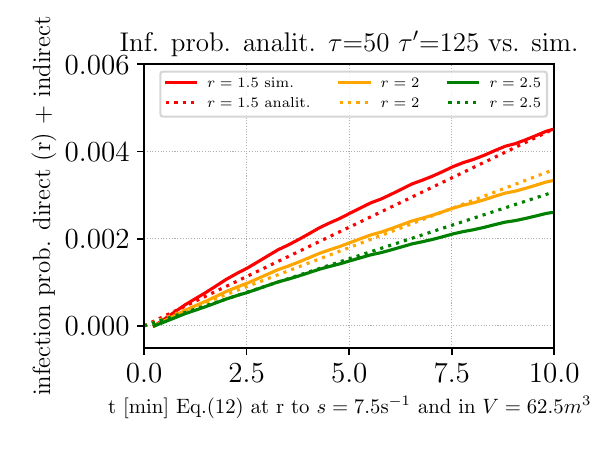}
	\includegraphics[width=0.45\textwidth]{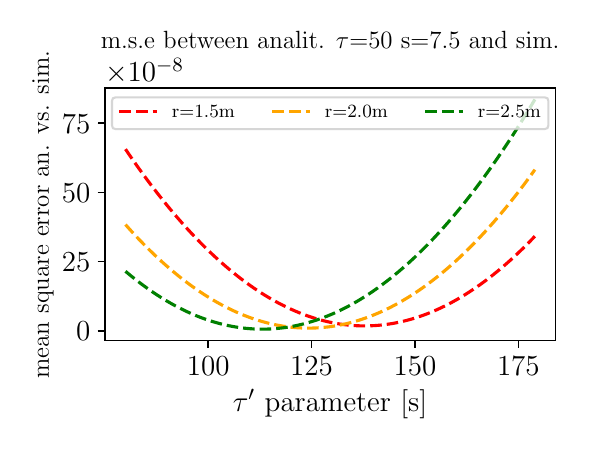}
	\caption{Absorb setting. The left panel presents the probability of infection computed from the analytical model and simulations as the function of the time the agent inhales contaminated air ($t$ [minutes]). The right panel presents the mean square error (m.s.e.) between the analytical model and simulations for various values of the $\tau'$ parameter. For both the analytical model and simulations, the probability of infection is computed via. Eq.~\eqref{eq::probability_air}, with immunity factor $N_b = 100$ from the number of inhaled droplets computed via simulations (solid) and analytical model (dashed). In the analytical model, the number of inhaled droplets is computed from Eq.~\eqref{eq::n_droplets} via droplet's density from proximity (direct mode) in  Eq.~\eqref{eq::crs} and environmental air (indirect mode) in Eq.~\eqref{eq::cbackgropud}.}\label{fig::simulations_inf}
\end{figure}

In Fig.~\ref{fig::simulations_inf}, we compare the probability of a healthy individual becoming infected due to proximity exposure and from contaminated environmental air. These probabilities, calculated using the analytical model, are compared with analogous probabilities obtained from simulations under the absorb setting. For the analytical model, we used a source rate of $s = 7.5$ s$^{-1}$ and a range of $\tau'$ values, which are significantly larger than the decay constant $\tau$ identified in Fig.~\ref{fig::simulations}.
Regarding proximity to the infected individual, we focus on distances of at least $1.5$ m, as indicated in the validity range of Fig.~\ref{fig::simulations}. Additionally, in the analytical model, we now incorporate the effect of the indirect infection pathway through contaminated air within the compartment. 
From Fig.~\ref{fig::simulations_inf}, we observe that the infection probabilities computed from the analytical model closely match those from simulations across different inhalation durations ($t$) of the contaminated air. Here, the inhalation time ($t$) is assumed to be significantly shorter than the duration of the emission stage, minimizing potential effects associated with the temporal separation of emission and absorption.
In Fig.~\ref{fig::simulations_inf} (right panel), we identify a reasonable range for the $\tau'$ parameter to be between $100$ and $150$ seconds, with a midpoint at $\tau' = 125$ seconds. This value of $\tau'$ is notably higher than the decay constant $\tau$ used for infection via direct contact, which aligns with expectations for indirect transmission in a larger air volume.

\begin{figure}[ht]
    \centering
\includegraphics[width=0.6\textwidth]{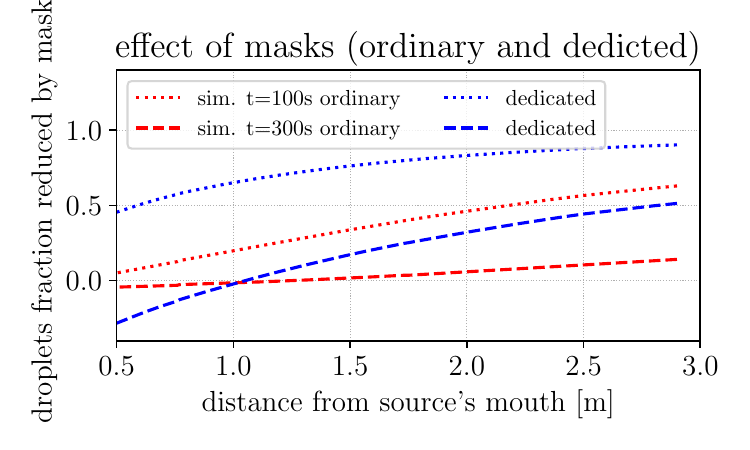}
	\caption{Simulation results of the reduction of droplet density due to application of various masks, compare the effect of the ordinary and dedicated masks. Simulation times higher than $300$s are not considered, as such simulation times would be much higher than the possible value of the decay constant $\tau$.}\label{fig::masks}
\end{figure}

Lastly, Fig.~\ref{fig::masks} presents the impact of various types of masks in the emit setting, showing the fraction of droplets reduced by mask usage, calculated as $\frac{c_{\text{no mask}}-c_\text{mask}}{c_{\text{no mask}}}$. This reduction factor is dependent on both distance and mask type. For instance, an effective mask can be modeled by reducing the source rate $s$ by approximately $40\%$, as suggested in \cite{leung2020respiratory}. An ordinary mask would yield a reduction in source rate of roughly $20\%$.

%%%%%%%%%%%%%%%%%%%%%%%%%%%%%%%%%%%%%%%%%%%%%%%%%%
\section{Agent-based simulations}\label{sec:abm-results}
%%%%%%%%%%%%%%%%%%%%%%%%%%%%%%%%%%%%%%%%%%%%%%%%%%

To bring our findings closer to industrial settings and demonstrate that a simple random walk model can simulate pathogen spread, we analyze disease transmission within a simplified simulated environment. Specifically, the goal of these simulations is to evaluate the relative impact of two primary transmission modes, \emph{direct} and \emph{indirect}, under varying model parameters. Using agent-based simulations, we compare the influence of the most probable \emph{direct} transmission from \emph{proximity to an infected individual} with \emph{indirect} transmission via \emph{contact with a contaminated surface}. We hypothesize that agent density, rather than the actual size of the industrial facility, plays a crucial role in determining infection spread. Additionally, for large industrial facilities, while the probability of infection via \emph{contaminated environmental air} may be negligible, worker contact with potentially contaminated tools and machinery remains a significant factor.

Technically, we employ an agent-based model designed to mimic the movement of individuals in an environment similar to an industrial workplace. We represent this environment as rectangular areas with a spatial resolution of $r = 1.5$ meters. Each simulation step corresponds to $\delta t = 60$ seconds ($1$ minute of real time), meaning that an 8-hour work shift equates to 480 simulation steps.

In many infections, such as COVID-19, infected individuals are not immediately infectious. To account for this, we incorporate a \emph{latency factor} wherein newly infected individuals do not transmit the infection to others. For this experiment, we set the latency period to one day, meaning that individuals newly infected within the simulation do not contribute to further transmission during the results presented in this section.

\begin{figure}[ht!]
\includegraphics[width=\textwidth]{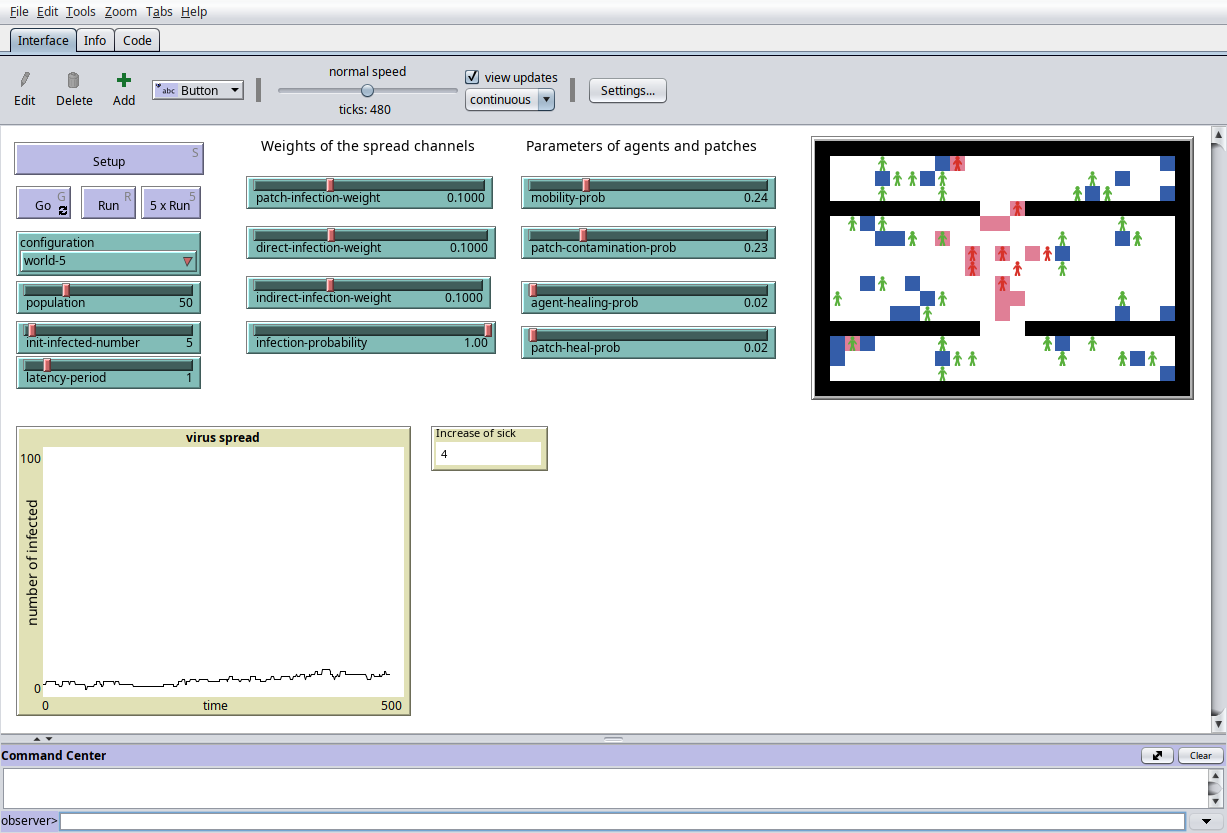}
\caption{User interface of the NetLogo model used in this paper. The screen presents an example of one run of the simulation cycle corresponding to one day. Black elements represent walls, blue elements represent machines or workplaces. The source code of this mode can be obtained from~\cite{indoors_virus_propagation_model}.}
\label{fig:netlogo-model-view}
\end{figure}

The elements within the simulated environment are designed to mimic compartments, machines, or workstations, as well as pathways between compartments, as illustrated in Fig.~\ref{fig:netlogo-model-view}. Agents move randomly within this space, with their movement controlled by a mobility parameter $\mu$, defined as the probability of changing position at each simulation step. To simulate the effect of working at a specific location, the random movement of agents is slowed near designated areas that represent workplaces. This is achieved by reducing the mobility factor $\mu$ by $10^{-2}$ in proximity to these workplace locations.The model used in this section was implemented using NetLogo multi-agent programmable modeling environment~\cite{wilensky1999netlogo}. 

At the start of the simulation, a predefined number of agents are designated as initially infected. Infected agents can transmit the infection to others through direct contact or indirectly by contaminating surfaces (or "patches") representing workplaces. Since the progression of the disease is assumed to be significantly longer than a single work shift, we assume that contaminated patches and infected agents do not undergo any recovery or decontamination during the shift. Based on these assumptions, we consider the following potential infection pathways:
\begin{enumerate}
\item \emph{Direct contact with an infected individual}. In our simulation, given its spatial resolution, an infection can potentially occur if two agents occupy the same pixel, corresponding to a distance of $r = 0.75$ m, or adjacent pixels at a distance of $r = 1.5$ m. The probability of infection during a $\delta t$ simulation step is estimated for small $\delta t$ using an approximation derived from Eq.~\eqref{eq::probability_air} as follows:
\begin{equation}
 \delta p_{\text{direct}} \approx a \frac{N(\delta t)}{N_b}.
 \end{equation}
Then using Eq.~\eqref{eq::n_droplets} and Eq.~\eqref{eq::crs} we have
 \begin{equation}
 \delta p_{\text{direct}} = a \frac{Q \hat{c}(r, s)}{N_b} \delta t.
 \end{equation}
Taking $Q = 19 $dm$^3$/minute = $3.16 \cdot 10^{-4} $m$^3$/s, $N_b = 100$, and $a = 0.1$, we compute  $\hat{c}(r, s)$ using Eq.~\eqref{eq::crs} with parameters $\tau = 50$s, $D = 0.05$m$^2$/s, $r_0 = 0.62$m, $d_{\max} = 2.6$m, and a source rate of $s = 66$ s$^{-1}$ which represents a coughing individual according to the discussion in Section~\ref{sec::analitical_model}. This setup is reasonably validated by the results in Fig.~\ref{fig::simulations} (left panel). We have:
%\begin{itemize}
\begin{equation}
    \delta p_{\text{direct}} \approx \begin{cases}
     0.0049$ minute${}^{-1}
    &\text{meet at the same patch} \\
     0.0013$ minute${}^{-1} &\text{meet at neighbor patches}
    \end{cases}
    \label{eq:sources}
\end{equation}
%\item not coughing but speaking - sources $s = 7.5$ s$^{-1}$:
%\begin{equation}
%    \delta p_{\text{direct}} \approx \begin{cases}
%    0.00054$ min${}^{-1}
%    &\text{meet at the same patch} \\
 %    0.00014$ min${}^{-1} &\text{meet at neighbour patch}
 %   \end{cases}
%\end{equation}
%\end{itemize}

\item \emph{Indirect infection through contact with a contaminated surface}. Since the probability of a patch becoming contaminated is highly dependent on various parameters (e.g., the size of the working area), we introduce a variable patch contamination probability to evaluate different scenarios. In our simulations, we assume that, at each $1$-minute step, a patch can become contaminated with a certain probability, referred to as the \emph{patch contamination probability}. This probability is varied from $0$ to $0.5$ with a specified resolution step to explore a range of possible contamination levels.

\begin{figure}[ht!]
\includegraphics{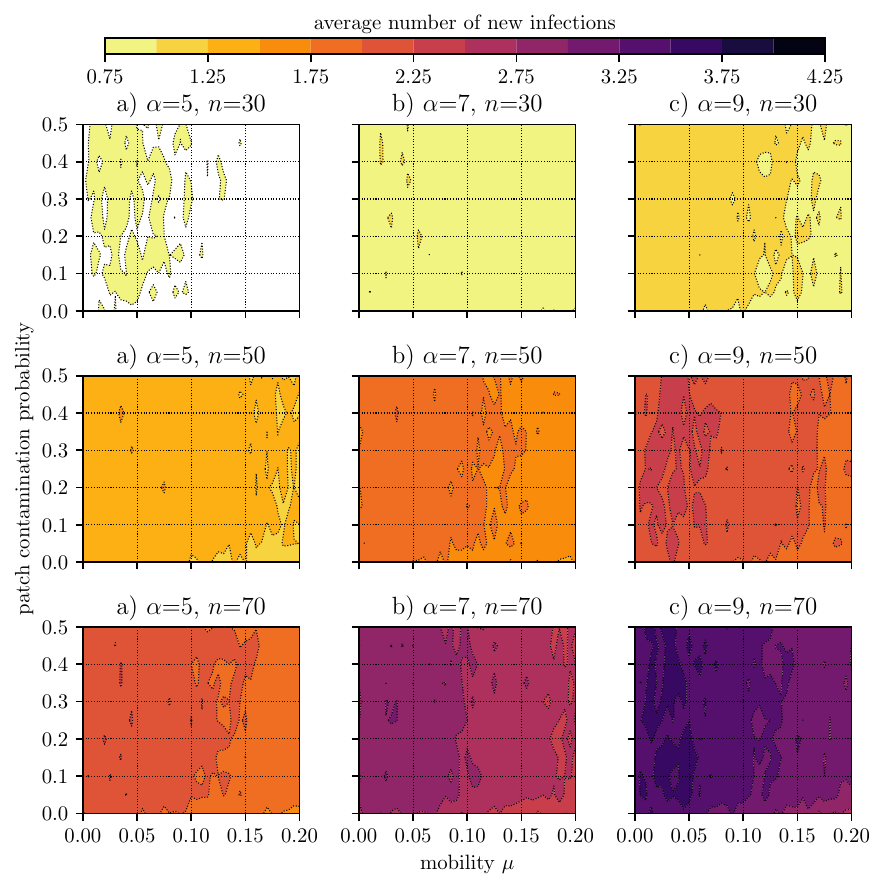}
\caption{Mean increase in the number of infected agents for different values of mobility parameter $\mu\in[0.0, 0.5]$, different populations, $n=30, 50$ and $70$ agents, and different number of initially infected agents, $\alpha=5, 7$ and $9$. Each plot represents the absolute increase in the number of agents infected during one day in the population $\alpha$ of initially infected agents, with different values of patch contamination probability, and with different mobility $\mu$ of the agents. We use the source rate of $s = 66$ s$^{-1}$ (see discussion on Eq.~\ref{eq:sources}). Mean is calculated over $10^{4}$ realizations. No healing process is considered.}
\label{fig:increase_large_s_varpop}
\end{figure}

To compute the probability of an agent getting infected from the surface, we use  Eq.~\eqref{eq::prob_surface} with $p_{\text{contact}} = 10^{-4}$ and $f_{\text{contact}} = 1$ minute$^{-1}$):
\begin{equation}
\delta p_{\text{undir. c}}  \approx 0.0001 \text{ minute}^{-1}.
\end{equation}
\end{enumerate}
 
Simulation results, showing the numbers of newly infected agents, are presented in Fig.~\ref{fig:increase_large_s_varpop}. The density of agents is changed between $n=30, 50$, and $70$.

The main finding is that higher concentrations of agents facilitate the spread of infection. Therefore, reducing the number of people working on-site simultaneously should be considered a primary defense strategy against infection spread. Additionally, we observe that while increased mobility generally leads to a reduction in new infections, there is a local maximum in infections at a mobility factor of $\mu \approx 0.05$. This indicates an important insight: there exists a specific level of mobility where the probability of infection is highest.

\begin{figure}[ht!]
\includegraphics{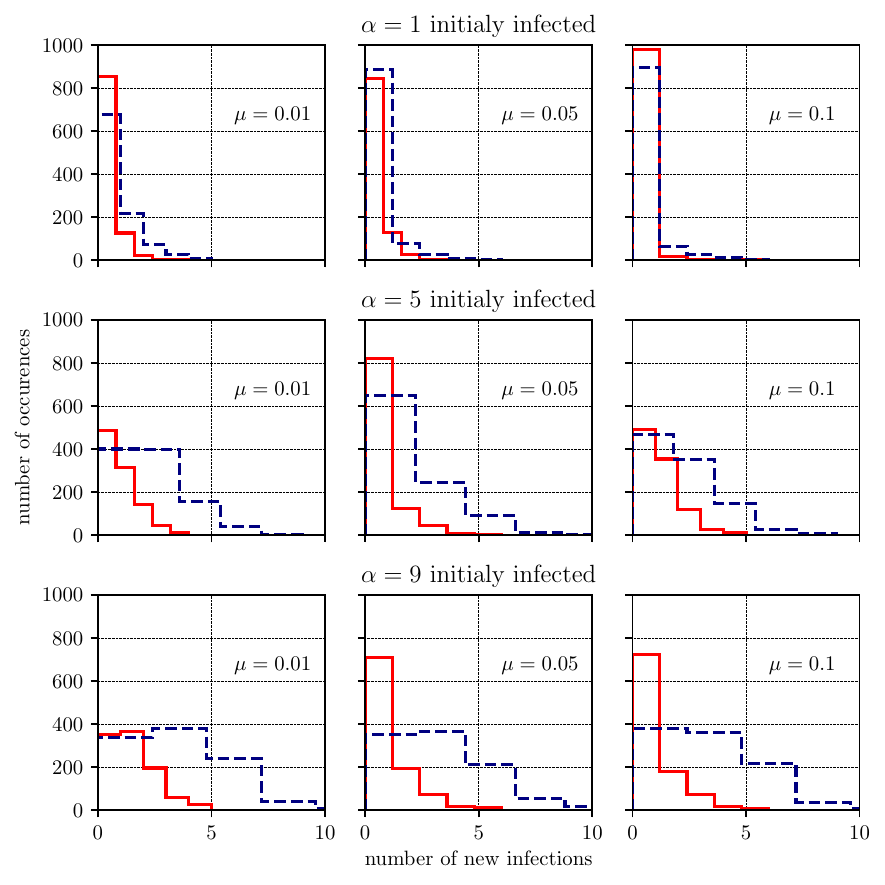}
\caption{Dependency between the number of infected agents and the agents' mobility $\mu$. Each row represents a series of histograms for different numbers of initially infected agents, while each column a series of histograms for different agents' mobility factor $\mu$. Histograms represent the occurrences (counts) of the number of new infections, for a fixed number of initially infected agents, for the case $n=30$ (red solid line) and the case $n=70$ (blue dashed line). Patch contamination probability is set to $0.5$. Histograms are calculated over $10^{4}$ realizations.}
\label{fig:exp6_pop100_world-1_hist}
\end{figure}

Another key aspect we examine is the impact of agent mobility and population density on infection spread. The results of this analysis are shown in Fig.~\ref{fig:exp6_pop100_world-1_hist}. In the top row of Fig.~\ref{fig:exp6_pop100_world-1_hist}, we observe that mobility does not significantly influence infection spread for a low number of initially infected agents. This effect holds for both small and large populations. When the initial proportion of infected agents is around 7\%-15\% of the population (middle row of Fig.~\ref{fig:exp6_pop100_world-1_hist}, with $\alpha=5$), an increase in mobility tends to reduce the number of new cases, regardless of population density. However, when the number of initially infected agents is higher, the density of all agents becomes critical to the dynamics of infection spread. In this scenario, as shown in the bottom row of Fig.~\ref{fig:exp6_pop100_world-1_hist}, higher agent density leads to a larger increase in new infections. Conversely, for a smaller population of $n=30$ agents -- and thus lower individual density -- the number of new infections decreases as mobility increases. This effect is not observed in the larger population of $n=70$ agents, indicating that, for larger populations, agent mobility has a less pronounced impact on infection rates, resulting in a more stable increase in the number of cases.

%%%%%%%%%%%%%%%%%%%%%%%%%%%%%%%%%%%%%%%%%%%%%%%%%%
\section{Conclusions}\label{sec::conclusions}
%%%%%%%%%%%%%%%%%%%%%%%%%%%%%%%%%%%%%%%%%%%%%%%%%%

In the wake of the COVID-19 pandemic, there is an increased awareness that outbreaks of dangerous pathogens are possible. Based on this experience, it has become clear that society must prepare for future epidemics (not necessarily COVID-19) in ways that minimize disruptions to the economy and supply chains. To aid in strategic planning for industrial operations during such outbreaks, our research provides several valuable insights.

We have developed a new analytical model of pathogen spread that is computationally efficient, making it a potentially favorable alternative to more resource-intensive simulations in some contexts. However, we acknowledge that this model relies on several simplifying assumptions and encompasses a wide range of parameters. A key finding is that the analytical model can approximate simulation results in many cases, though proper parameter calibration is essential. This can be achieved by comparing the analytical model's outcomes with simulation results.
In this example, we demonstrate how simulations can validate analytical models and help identify the parameter ranges in which the models are accurate. This approach allows simulations to be run once during the initial validation phase, after which the analytical model can be applied for day-to-day monitoring and decision-making to maintain safety in industrial settings. From a validation perspective, we also recognize the limitations of our analytical model, particularly when modeling infection spread near the source, as shown in Fig.~\ref{fig::simulations}.

To validate the analytical model from an additional perspective, we can outline some general countermeasures suggested by the model. These countermeasures include: removing workers who are coughing or sneezing from shared workspaces or requiring such individuals to wear high-quality masks immediately. Subsequently, additional measures could focus on reducing agent density and limiting mobility (or mandating high-quality masks for workers with high mobility). 
Countermeasures should initially prioritize the  \emph{direct transmission}. Measures targeting \emph{indirect transmission} routes, such as surface disinfection, can be implemented subsequently as supportive actions.

Finally, the role of agent-based simulations was demonstrative. We recognize that it would be valuable if the authors applied the simulation to a real-world industrial topology with realistic movements within the plant, perhaps based on actual measurements. Although the direct industrial data we have are confidential. Further more detailed research shall enhance the relevance and applicability of the agent-based model, which currently remains a largely theoretical exercise. %Nevertheless, we draw an unexpected conclusion, that there is a certain level of mobility of agents, where the infection is most probable. This conclusion needs further investigation.

%%%%%%%%%%%%%%%%%%%%%%%%%%%%%%%%%%%%%%%%%%%%%%%%%%

\section*{Data availability}
The source code used to simulate the model discussed in this work, along with sample simulation results, can be accessed at~\cite{indoors_virus_propagation_model}.

%\section*{Author contributions statement}
%A.S., J.M., K.D. - conceptualisation, A.S., J.M. - preparing simulations, A.S., J.M. - running simulations, K.D. - preparing analytical model, A.S., J.M., K.D. - data analysis, A.S., J.M., K.D. - manuscript writing. 

\section*{Declaration of competing interest}
The authors declare that they have no known competing financial interests or personal relationships that could have appeared to influence the work reported in this paper. All authors reviewed the manuscript.
%%%%%%%%%%%%%%%%%%%%%%%%%%%%%%%%%%%%%%%%%%%%%%%%%%

\section*{Acknowledgments} 
The authors would like to thank Ryszard Winiarczyk for motivating discussions and valuable tips on simulations of droplet spreading.

K.D. and A.S. acknowledge the support from The National Centre for Research and Development, Poland, project number: POIR.01.01.01-00-2612/20.

J.M. would like to acknowledge that his work has been motivated by personal curiosity and received no support from any agency.

\end{document}